\documentclass[11pt]{article}
\usepackage{bm}
\usepackage{amsmath}
\usepackage{latexsym}

\begin{document}

%-------------------------------------------------------------------------------------
%The following line was need since the .ps output from dvips seemed to have
%a headheight that was 1/2 inch too small. I increased it by 1/2 inch.
%If this pushes the headheight for your .ps file down an EXTRA 1/2 inch, then
%remove the following line.

\setlength{\headheight}{0.0in}  %original value 0.0; I found 0.5in works better: PMA
%------------------------------------------------------------------------------------

% You should use BibTeX and apsrev.bst for references
\bibliographystyle{apsrev}

%\preprint{October 10, 2000}

%===========================================================================
%                              Title of Paper
%===========================================================================

\title{The phase of a quantum mechanical particle in curved spacetime}
% Optional argument for running titles on pages
%\title[]{}



%===========================================================================
%                             Authors and affiliations
%===========================================================================

%==========================================================================
% Plain latex titlepage
%==========================================================================

\author{P.~M.~Alsing \footnote{Contact author for correspondences.}\\
\textit{Albuquerque High Performance Computing Center},\\
\textit{University of New Mexico, Albuquerque, NM, 87131} \\
\textit{alsing@ahpcc.unm.edu}
\\ \\
J.~C.~Evans \\
\textit{Department of Physics and Astronomy},\\
\textit{University of Puget Sound, Tacoma, WA, 98416}\\
\textit{jcevans@ups.edu}
\\ \\
K.~K.~Nandi \\
\textit{Department of Mathematics, University of North Bengal}, \\ 
\textit{Darjeeling (WB) 734430, India} \\
\textit{nbumath@dte.vsnl.net.in}}  %\email[]{kamalnandi@hotmail.com}

%==========================================================================
% Revtex4 titlepage
%==========================================================================

%\author{P.~M.~Alsing}
%\email[]{alsing@ahpcc.unm.edu}
%\affiliation{Albuquerque High Performance Computing Center,\\
%University of New Mexico, Albuquerque, NM, 87131}

%\author{J.~C.~Evans}
%\email[]{jcevans@ups.edu}
%\affiliation{Department of Physics and Astronomy,\\
%University of Puget Sound, Tacoma, WA, 98416}

%\author{K.~K.~Nandi}
%\email[]{kamalnandi@hotmail.com}
%\affiliation{Department of Mathematics, University of North Bengal, \\ 
%Darjeeling (WB) 734430, India}

% For some reason, adding the date made a blank second page. I had to add some
% negative space to prevent this. PMA
%\vspace*{-.5in}
\date{}
\maketitle  %this \maketitle is for the plain latex version
%-------------------------------------------------------------------------------------

%===========================================================================
%                            Abstract
%===========================================================================
\vspace*{-.35in}
\begin{abstract}
We investigate the quantum mechanical wave equations for  free particles of spin $0,1/2,1$ 
in the background of an arbitrary static gravitational field in order to explicitly
determine if the phase of the wavefunction is  $S/\hbar = \int p_{\mu}\,dx^{\mu} / \hbar$,
as is often quoted in the literature. We work in isotropic coordinates where
the wave equations have a simple managable form and do not make a weak
gravitational field approximation. We interpret these wave equations in terms
of a quantum mechanical particle moving in medium with a  spatially
varying effective index of refraction.
Due to the first order spatial derivative
structure of the Dirac equation in curved spacetime, only the spin $1/2$ particle has 
\textit{exactly} the quantum mechanical phase as indicated above. The second order 
spatial derivative structure
of the spin $0$ and spin $1$ wave equations yield the above phase only to
lowest order in $\hbar$. We develop a WKB approximation for the solution of the
spin $0$ and spin $1$ wave equations and explore amplitude and
phase corrections beyond the lowest order in $\hbar$. 
For the spin $1/2$ particle we calculate the phase appropriate for 
neutrino flavor oscillations.
\end{abstract}
%\newpage

% Quantum Mechanics, Quantum statistical mechanics, Quantum transport,
% Brownian motion
%\pacs{}

%\maketitle must follow title, authors, abstract and \pacs
%\maketitle %this is for REVTEX4 version

%===========================================================================
%                    Body of Paper here 
%===========================================================================
%  - Use proper section commands
%  - References should be done using the \cite, \ref, and \label commands

\section{Introduction}
The phase of a quantum mechanical particle in curved spacetime has been of
considerable interest both theoretically and experimentally for many years.
In the late 1970's  researchers were interested in explaining
the quantum mechanical interference fringes for neutrons traversing different
paths in the Earth's gravitational field \cite{Gre79}. Recently, there has been
renewed interest in this topic in order to investigate the interplay between
gravitation and the quantum mechanical principle of linear superposition in
relation to flavor oscillations of neutrinos in the context of the
solar neutrino anomaly, and type-II supernova \cite{For97,Ahl98,Cap99}.

Most computations of the quantum mechanical phase (QMP) of a free particle in curved
spacetime refer back to the seminal article by Stodolsky \cite{Sto79} who argued
that the relativistically invariant phase $S/\hbar$  would be given by
\begin{equation}
\label{eq1}
S(\bm{r},t) / \hbar = \frac{1}{\hbar}\,\int_{\bm{r}_A,\,t_A}^{\bm{r}_B,\,t_B} p_{\mu}\,dx^{\mu}. 
\end{equation}
In Eq.~(\ref{eq1}) $p_{\mu} = m g_{\mu\nu} u^{\nu}$ is the general relativistic
4-momentum, and $u^{\mu} = dx^{\mu}/d\tau$ is the 4-velocity such that
$g_{\mu\nu} \, u^{\mu} \, u^{\nu}=1$. The phase is computed along a geodesic
connecting the points $(\bm{r}_A,\,t_A)$ and $(\bm{r}_B,\,t_B)$. 
The quantum mechanical particle is assumed to be a test particle in the sense
that it moves in the background of the general relativistic metric and does
not generate it own gravitational field.
For any arbitrary metric in general relativity (GR), static or otherwise, 
we have a relationship amongst the momenta, called the \textit{mass shell constraint}, given by
\begin{equation}
\label{eq2}
g^{\mu\,\nu} \, p_{\mu} \, p_{\nu} = (m c_0)^2,
\end{equation}
where $m$ is the rest mass of the particle and $c_0$ is the vacuum speed of light.
The form of the scalar quantum mechanical
wave function proposed by Stodolsky was  
\begin{equation}
\label{eq3}
\psi(\bm{r},t) = A \, e^{i\,S(\bm{r},t)/\hbar},
\end{equation}
where the amplitude $A$ is assumed constant.

It is clear from  Eq.~(\ref{eq3}) that a single spatial derivative of $\psi(\bm{r},t)$
can generate $p_{\mu}$, but that two spatial derivatives cannot generate  the 
mass shell constraint Eq.~(\ref{eq2}),
since imaginary cross terms involving gradients of $p_{\mu}$ will be produced.
Therefore, for second order wave equations, for which  spin $0$ and spin $1$ particles are particular
cases, the QMP cannot \textit{exactly} take the form of  Eq.~(\ref{eq1}). In addition, the
form of $\psi(\bm{r},t)$ in  Eq.~(\ref{eq3}) must involve an amplitude change, which is well
known from the standard WKB approximation of the wave function in non-relativistic 
quantum mechanics \cite{Sch68}.

In this paper we examine the general relativistic wave equations for  quantum mechanical 
particles of spin $0,1/2,1$ in the background of a generic static metric
of arbitrary strength which can be written in isotropic form. 
We do not make a weak gravitational field approximation, which is often 
the case in the literature \cite{Don86}.
By working in isotropic coordinates, the wave equations are expressed in simple forms,
which are then easily intrepreted. 
We explicitly demonstrate that for the Dirac equation in curved spacetime the form of the
QMP is given exactly by  Eq.~(\ref{eq1}), and that  Eq.~(\ref{eq2}) is exactly satisfied.
This result is directly traceable to the first order spatial 
derivative structure of the Dirac equation.
The only difference from Stodolsky's proposal is that amplitude in  Eq.~(\ref{eq3}) takes on a simple 
spatially varying form. For particles of spin $0$ and spin $1$, the wave equation is of
second order in both time and space. We develop a WKB approximation for  $\psi(\bm{r},t)$
which to lowest order in $\hbar$ has the form of  Eq.~(\ref{eq1}) and satisfies Eq.~(\ref{eq2}).
We formally solve for $\psi$ to all orders in $\hbar$. However, since the WKB expansion is
only an asymptotic series \cite{Ben78} we only examine the next higher order phase and amplitude
corrections. We give an interpretation of the above wave equations in terms of a particle moving
in a medium with a spatially varying index of refraction  $n(\bm{r})$ as discussed in
Evans \textit{ et. al.} \cite{Eva96} and Alsing \cite{Als98} (see also \cite{Eva00}).

The outline of the paper is as follows. 
In Section \ref{sec2} we briefly review Stodolsky's reasoning for the form of the phase of
    a quantum mechanical particle in curved spacetime as given in  Eq.~(\ref{eq1}).
In Section \ref{sec3} we examine the general form of the quantum mechanical 
   wave equation for a free particle in a background curved spacetime. 
In Section  \ref{sec4} we examine the Dirac equation in curved spacetime and 
   show that in isotropic coordinates, it takes on a simple form even for arbitrary 
   strength gravitational fields.
In Section \ref{sec5} we examine the wave equation for a scalar particle 
   in curved spacetime and develop
   a WKB solution. We develop $\hbar$-dependent quantum corrections  beyond the 
   lowest order WKB amplitude and phase approximations found in standard textbooks. 
   This WKB approximation has applicability to the curved spacetime spin $1$ wave equation 
   since the latter has 
   the same form as the scalar wave equation for each of the vector components of the 
   wave function.
   Throughout the discussion
   we draw the analogy to a particle moving in a medium with a spatially varying index of refraction,
   which the form of these wave equations explicitly exhibit.
In Section \ref{sec6} we relate the classical optical-mechanical analogy 
   of the lowest order approximation to the QMP.  
In Section \ref{sec7} we calculate the  QMP for a  spin $1/2$ particle 
   appropriate for neutrino flavor oscillations in a gravitational field and
   relate it to the effective index of refraction discussed in the previous section.
In Section \ref{sec8} we summarize our results and discuss their implications in light of current
   calculations in the literature.

\section{Stodolsky's proposal for the  phase of a quantum mechanical particle in curved spacetime}
\label{sec2}
In 1979, Stodolsky \cite{Sto79} argued the phase of a spinless quantum mechanical 
particle should take the
form of Eq.~(\ref{eq1}). In the following we summarize his reasoning. In flat Minkowski spacetime
elementary quantum mechanics tells us the phase of the particle's wavefunction is given by dimensionless
quantity $(\bm{p}\cdot\bm{x} - E\,t)/\hbar$. The numerator of this expression has units
of action,  $m c_0 \times$ (proper distance). Since the particle is taken to be
a test  particle, and therefore does not generate its own gravitational field, classically it
would follow a geodesic. This is assumed to be true quantum mechanically as well,
independent of the particle's spin. For an arbitrary metric 
$ds^2 = g_{\mu\nu} \, dx^{\mu} \, dx^{\nu}$, Stodolsky proposed the phase $\Phi$ accumulated 
by the particle in traversing the geodesic connecting the spacetime points $A$ and $B$  to be 
\begin{equation}
\label{eq4}
\Phi = \frac{c_0}{\hbar} \, \int_A^B m \, ds \equiv S/\hbar.   
\end{equation}
The particle's quantum mechanical wavefunction is then taken to be proportional to
the phase factor  $e^{i\,\Phi}$.
The integral appearing in Eq.~(\ref{eq4}) is just the relativistic action for
a particle moving on a geodesic \cite{Lau65}.
Stodolsky states (\cite{Sto79}, p392), ``If we wish the phase to be an invariant
and to agree with elementary quantum mechanics this seems to be the only choice.''
Furthermore, dividing the metric by $ds$ and defining 
$p_{\mu} = m c_0 \, g_{\mu\nu} \, dx^{\nu}/ds$ as the canonical general relativistic momentum 
yields  Eq.~(\ref{eq1}). Stodolsky finds additional reassurance  
from the observation that  Eq.~(\ref{eq1})
also appears reasonable from the Feynman path integral approach to quantum mechanics.
If  Eq.~(\ref{eq4}) is valid for paths neighboring the classical path then one obtains
in the usual way that the actual classical path is the one for which $\delta\Phi=0$.
But from the form of Eq.~(\ref{eq4}), this is just the classical relativistic condition for a geodesic.
In his paper \cite{Sto79} Stodolsky examines  Eq.~(\ref{eq1}) in the limit of weak
gravitational fields for a for both static and stationary metrics. 

%%%%%%%%%%%%%%%%%%%%%%%%%%%%%%%%%%%%%%%%%%%%%%%%%%%%%%%%%%%%%%%%%%%%%%%%%%%%%%%%%%%%%%%%%%%%%%%
The purpose of this work is to explicitly test whether Eq.~(\ref{eq1}) 
is a valid solution of the
general relativistic wave equations for particles of spin $0, 1/2, 1$.
To this effect, we consider an arbitrary static gravitational field
which can be written in isotropic coordinates \cite{Adl75}. 
For such gravitational fields, the spatial portion of the
metric can be written in a conformally flat form,
\begin{equation}
\label{eq6}
ds^2 = \Omega^2(\bm{r}) \, c_0^2 \, dt^2 - \frac{1}{\Phi^2(\bm{r})} \, ( dx^2 + dy^2 + dz^2) \equiv
     \Omega^2(\bm{r}) \, c_0^2 \, dt^2 - \frac{1}{\Phi^2(\bm{r})}\,|d\bm{r}|^2.
% \frac{|d\bm{r}|^2}{\Phi^2(\bm{r})}, 
\end{equation}
In the above, $r$ is the isotropic radial marker. 

Many static gravitational fields of physical
interest can be written in the form of Eq.~(\ref{eq6}). The important class
of static, spherically symmetric metrics written in the usual spherical coordinates
$\{t,r',\theta,\phi \}$ have the general, non-isotropic  form 
\begin{equation}
\label{eq5}
ds^2 = \frac{F(r')}{G(r')} \, c_0^2 \, dt^2 - \frac{dr^{2}}{F(r')} 
     - R^2(r') \, (\, d\theta^2 + sin^2(\theta) \, d\phi^2 \,).  
\end{equation}
For example, the Schwarzschild metric has $F(r') = ( 1 - r'_s/ r')$, $G(r') = 1$, and 
$R(r') = r'$ where $r'$ is the usual radial coordinate measured by an observer
at infinity and $r'_s=2 G M /c_0^2$ is the Schwarzschild radius. 
After transforming the Schwarzschild metric
to isotropic coordinates $\{t,r,\theta,\phi \}$ one obtains
$\Omega(r) = (1+r_s/r)/ (1-r_s/r)$ and 
$\Phi(r) = (1 + r_s/r)^{-2}$ where $r_s = r'_s/2$. 
The relationship between the coordinate radius $r'$ and
the isotropic radius $r$ is given by $r' = r \, \Phi^{-1}(r) = r\,(1+r_s/r)^2$.
Other interesting static spherically symmetric metrics for both
Einstein gravitation theory and non-Einstein gravity theories  can
be found in \cite{Nan00b}.
%%%%%%%%%%%%%%%%%%%%%%%%%%%%%%%%%%%%%%%%%%%%%%%%%%%%%%%%%%%%%%%%%%%%%%%%%%%%%%%%%%%%%%%%%%%%%%%

The utility of isotropic coordinates stems from the ease by which we can define
an effective index of refraction for the gravitational field. By setting $ds=0$ in
Eq.~(\ref{eq6}) we obtain the coordinate speed of light $c(\bm{r})$ as
\begin{subequations}
\begin{eqnarray}
c(\bm{r}) &=& \left|\frac{d\bm{r}}{dt}\right| = c_0 \, \Omega(\bm{r}) \, \Phi(\bm{r}) \equiv \frac{c_0}{n(\bm{r})}, \label{eq7a} \\
n(\bm{r}) &\equiv& \frac{1}{ \Omega(\bm{r}) \, \Phi(\bm{r})}. \label{eq7b}
\end{eqnarray}
\end{subequations}
The paths of both massive and massless particles in such metrics (i.e. geodesics) can be
interpreted as motion through a medium of index with an effective 
refraction $n(\bm{r})$. For the Schwarzschild metric
we have $n(\bm{r}) = (1+r_s/r)^3/(1-r_s/r)$. Evans \textit{et al.} \cite{Eva96} used this
concept to write the geodesic equations of 
motion for static metrics in Newtonian ``F=ma'' form. 
Alsing \cite{Als98} later extended this idea to the case of stationary metrics.
In this paper, the quantity $n(\bm{r})$ will continually arise in the wave equations developed, 
and will be naturally interpreted as an effective index of refraction.

For the metric given by  Eq.~(\ref{eq6}) we can write the mass shell constraint  Eq.~(\ref{eq2}) as
\begin{equation}
\label{eq8}
 E^2 =  \left(\frac{\bm{p}(\bm{r})c_0}{n(\bm{r})}\right)^2 + \left(\Omega(\bm{r})\,m\,c_0^2\right)^2,
\end{equation}
where we define the 4-momentum as $p^{\mu} = (E/c_0,\,\bm{p})$.
This expression reduces to the ordinary flat space, special relativistic form 
$E^2 = (\bm{p}c_0)^2 + (m\,c_0^2)^2$ as $r\to\infty$. This naturally leads us to
propose the quantization rules
\begin{equation}
 \label{eq9}  E = \hbar\,\omega, \qquad \bm{p} = \hbar\,\bm{k},
\end{equation}
which define the frequency $\omega$ and wave vector $\bm{k}$.
Using Eq.~(\ref{eq9}) in  Eq.~(\ref{eq8}), we can rewrite the mass shell constraint as
\begin{equation}
\label{eq10}
n^2(\bm{r})\,k_0^2 + k^2(\bm{r}) = k_c^2(\bm{r}).
\end{equation}
In Eq.~(\ref{eq10}) we have defined the 4-wave vector as $k^{\mu} = (k_0,\bm{k})$
with $k_0=\omega/c_0$, $k^2 =\bm{k}\cdot\bm{k}$ 
and $k_c(\bm{r})\equiv (\lambda_c\Phi(\bm{r}))^{-1}$ where 
$\lambda_c= \hbar/(m\,c_0)$ is the usual Compton wavelength for a particle of mass $m$. 
Note the wave vector $\bm{k}(\bm{r})$ and hence the velocity of the particle is 
position dependent as appropriate for a particle in a medium with index of refraction $n(\bm{r})$.
%Although these quantization rules are not essential for the
%discussions in the rest of the paper (i.e. we could express all formulas in terms of the classical
%variables $E$ and $\bm{p}$) they nonetheless are convenient and natural definitions, and 
%so we put them forth.  

\section{Quantum wave equations in curved spacetime}
\label{sec3}
In practice, we only know
how to quantize wave equations in flat Minkowski space \cite{note1}.
Therefore, a natural way to describe curved spacetime is to erect local coordinate axes named
\textit{vierbeins} or \textit{tetrads} \cite{Bir82,Mis73,Wei72} at each point 
$X$ in spacetime and then project all tensor quantities onto these local, 
Lorentzian inertial frame axes. At each point $X$ the local metric
takes the flat spacetime form, $\eta_{ab}=\,$ diagonal~$\{1,-1,-1,-1\}$. Throughout this paper,
Latin indices $\{a,b,c,\ldots\}$ near the beginning of the alphabet
will refer to the local inertial frame with values $\{0,1,2,3\}$
while Greek indices $\{\mu,\nu,\lambda,\ldots\}$ will refer
to the general coordinate system $x^{\mu}$ with values $\{0,1,2,3\}$. 
The tetrads  $e_{\mu}^{\hspace{.5em}a}(x)$ and $e_a^{\hspace{.5em}\mu}(x)$ are defined by
\begin{equation}
 \label{eq11}
  g_{\mu\nu}(x) = e_{\mu}^{\hspace{.5em}a}(x) \, e_{\nu}^{\hspace{.5em}b}(x) \, \eta_{ab}, \qquad  
  g^{\mu\nu}(x) = e_a^{\hspace{.5em}\mu}(x) \, e_b^{\hspace{.5em}\nu}(x) \, \eta^{ab}.
\end{equation}
Contravariant and covariant vectors $V^{\mu}$ and $V_{\mu}$ in the
general coordinate system can be expressed
as vectors $V^{a}$ and $V_{a}$ in the local inertial frame 
(and visa versa)
by means of the transformations
\begin{eqnarray*}
V^{\mu}(x) = V^{a}(x) \, e_{a}^{\hspace{.5em}\mu}(x), &\qquad& 
V^{a}(x) = V^{\mu}(x) \, e_{\mu}^{\hspace{.5em}a}(x), \\
V_{\mu}(x) = e_{\mu}^{\hspace{.5em}a}(x)\,V_{a}(x), &\qquad& 
V_{a}(x) = e_{a}^{\hspace{.5em}\mu}(x)\,V_{\mu}(x).
\end{eqnarray*}
Local inertial frame indices are raised and lowered with the flat spacetime metric
$\eta_{ab}$, while all general coordinate frame indices are raised and lowered with
the metric $g_{\mu\nu}$.

The prescription for generalizing a flat spacetime wave equation to curved spacetime 
proceeds as follows: 
(i) begin with the appropriate flat spacetime Lagrangian for the wave equation of interest, 
(ii) replace all local partial coordinate derivatives by covariant derivatives via
$\partial_a \to e_a^{\hspace{.5em}\mu}(x)\,\nabla_{\mu}$, and (iii) contract
all vectors, tensors, etc. into $n$-biens, 
($V^{a}(x) \to V^{\mu}(x)\,e_{\mu}^{\hspace{.5em}a}(x)\,$, etc.). 

For a field $\psi(x)$ of arbitrary spin the spin covariant derivative $\nabla_{\mu}$ is
defined by \cite{Bir82,Wei72,Cha76,Law90}
\begin{subequations}
\label{eq12}
\begin{eqnarray}
\nabla_{\nu}\,\psi(x) &=& [\,\partial_{\nu} + \Omega_{\nu}(x)\,] \, \psi(x),  \label{eq12a}\\
\Omega_{\nu}(x) &\equiv& -\frac{i}{4}\,\omega_{ab\nu}(x)\,\sigma^{ab} = 
                         \frac{1}{8}\,\omega_{ab\nu}(x)\,[\,\gamma^a,\,\gamma^b\,], 
                         \quad\hspace{-1em} \textrm{spin 1/2}\label{eq12b} \\
\Omega_{\nu}(x) &\equiv& \frac{1}{2}\,\omega_{ab\nu}(x)\,\Sigma^{ab}.
                          \quad\hspace{8em}\textrm{arbitrary spin}\label{eq12c} 
\end{eqnarray}
\end{subequations}
In Eq.~(\ref{eq12}) the Fock-Ivanenko coefficients $\Omega_{\nu}(x)$
(not to be confused with $\Omega(\bm{r})\equiv\sqrt{g_{00}(\bm{r})}$ in the isotropic metric Eq.~(\ref{eq6}))
are defined in terms of the spin connection coefficients $\omega^{a}_{\hspace{.5em}b\nu}$ given by
\begin{subequations}
\begin{eqnarray}
\label{eq12.5}
\omega^{a}_{\hspace{.5em}b\nu} &=& e_{\mu}^{\hspace{.5em}a}\,\left(e_{b}^{\hspace{.5em}\mu}\right)_{;\nu}
=e_{\mu}^{\hspace{.5em}a}\,  \left( \partial_{\nu}\,e^{\hspace{.5em}\mu}_{b} 
                    + \, e_b^{\hspace{.5em}\sigma} \, 
                      \Gamma^{\mu}_{\sigma\nu} \right),  \label{eq12.5a}\\ 
\omega_{ab\nu} &=& e^{\mu}_{\hspace{.5em}a} \, e_{b\mu;\nu},  
              \qquad \left( e^{\mu}_{\hspace{.5em}a} \equiv g^{\mu\nu}\,\eta_{ab} \,e_{\nu}^{\hspace{.5em}b}   
                     \right), \label{eq12.5b} 
\end{eqnarray}
\end{subequations}
which are antisymmetric in their Lorentzian indices $\omega_{ba\nu} = -\omega_{ab\nu}$.
The semicolon denotes the usual Riemannian covariant derivative,
$V^{\mu}_{\hspace{.5em};\nu} \equiv \partial_{\nu} V^{\mu} + 
\Gamma^{\mu}_{\hspace{.5em}\sigma\nu} V^{\sigma}$. 
Since $\eta_{ab;\sigma} = 0$ and the Riemannian metric
compatibility condition gives $g_{\mu\nu;\sigma}=0$, we can freely raise and lower both Lorentzian (Latin)
and general coordinate indices (Greek) within a Riemannian covariant derivative operation.
Note  Eq.~(\ref{eq12.5a}) can be rearranged to define the action of the
spin covariant derivative $\nabla_{\nu}$ on the tetrad $e_{a}^{\hspace{.5em}\mu}$,
\begin{equation}
\label{12.75}
\nabla_{\nu}\,e_{a}^{\hspace{.5em}\mu} \equiv \partial_{\nu}\,e_{a}^{\hspace{.5em}\mu} 
                                  +  \Gamma^{\mu}_{\sigma\nu} \, e_{a}^{\hspace{.5em}\sigma} 
                                  -  \omega^{b}_{\hspace{.5em}a\nu}\,e_{b}^{\hspace{.5em}\mu} = 0,
\end{equation} 
analogous to the Riemannian metric compatibility condition.
Mnemonically, for the spin covariant derivative of a quantity 
$T^{ab\ldots\mu\nu\ldots}_{cd\ldots\gamma\delta\ldots}$ 
with a mixed set of index types,
each general coordinate index receives
a contribution from the metric connection and each local Lorentzian index receives 
a contribution from the spin connection.

Finally, in the last expression Eq.~(\ref{eq12c}) the constant matrices $\Sigma^{ab}$ 
are the generators of the Lorentz group for an arbitrary value of the spin  \cite{Bir82,Wei72}.
For the specific case of spin $1/2$,  $\Sigma^{ab} = 1/4 [\,\gamma^a,\,\gamma^b\,]$,
where $\gamma^a$ are the constant Lorentzian gamma matrices, with an explicit representation  
given in Appendix \ref{appendixA}. In addition, 
we have used the conventional notation $\sigma^{ab} = i/2\,[\,\gamma^a,\,\gamma^b\,]$
in  Eq.~(\ref{eq12b}).

We are now ready to generalize the flat spacetime Lagrangian  $\mathcal{L}(x)$ for the 
fields of interest to curved spacetime. Since in flat spacetime the action is given by
$S_{flat} = \int\,d^4 x \mathcal{L}(x)$, the curved spacetime analogue 
of $\mathcal{L}(x)$ must involve the volume factor 
$\sqrt{-g(x)}=\textrm{det}\Bigl(e_{\mu}^{\hspace{.5em}a}(x)\Bigr)$,
where $g(x)\equiv\textrm{det}\Bigl(g_{\mu\nu}(x)\Bigr)$, 
in order that  $S_{curved}$ transforms as a scalar.

\subsection{Spin 0 field}
\label{sec3.0}
For a spin $0$ field $\phi(x)$,  $\Sigma^{ab} = 0$, 
$\partial_a \phi(x) \to e_a^{\hspace{.5em}\mu}\nabla_{\mu} \phi(x)$ with
$\nabla_{\mu}\phi(x) = \partial_{\mu}\phi(x)$, and we have
\begin{eqnarray}
\label{eq13}
\mathcal{L}(x) &=& \frac{1}{2}\, \left( \eta^{ab} \, \partial_a \, \phi \, \partial_b \, \phi 
                 - \frac{m^2\,c_0^2}{\hbar^2} \phi \right) 
            \nonumber \\
            &\to& \frac{1}{2}\,(-g)^{1/2}  
                  \left(\,\eta^{ab} e_a^{\hspace{.5em}\mu} \, \partial_{\mu} \, \phi \, 
                              e_b^{\hspace{.5em}\nu} \, \partial_{\nu} \, \phi 
                            - \frac{m^2\,c_0^2}{\hbar^2}\,\phi \,\right). 
\end{eqnarray}
Variation of the Lagrangian with respect to $\phi(x)$ yields the general relativistic Klein-Gordon
equation (GRKGE)
\begin{eqnarray}
 \label{eq14}
&\mbox{}& \hspace{-3.5em}
\left(\, \Box + \frac{m^2 \, c_0^2}{\hbar^2}  \,\right) \phi(x) \equiv
 g^{\mu\nu}\,\phi_{;\mu;\nu} +  (m c_0/\hbar)^2 \phi(x) 
%\left(\, g^{\mu\nu}\,\nabla_{\mu} \nabla_{\nu} + \frac{m^2 \, c_0^2}{\hbar^2}  \,\right) \phi(x) 
\nonumber \\
&=& 
\left(g^{\mu\nu}\partial_{\mu}\partial_{\nu} - g^{\mu\nu}\Gamma^{\lambda}_{\mu\nu}\partial_{\lambda} 
+ \frac{m^2 \, c_0^2}{\hbar^2} %+ (m c_0/\hbar)^2 
\right)\phi(x) =0,
\end{eqnarray}
where we have defined the covariant Laplace-Beltrami operator 
$\Box\,\phi(x) \equiv g^{\mu\nu}\,\phi_{;\mu;\nu}$.
For the case of spin $0$ we can actually derive Eq.~(\ref{eq14}) much more directly from 
Eq.~(\ref{eq2}). Classically, the Lagrangian for geodesic motion is given by 
$L = m\,c_0\,ds =  m\,c_0\,\sqrt{ g_{\mu\nu}\,\dot{x}^{\mu}\,\dot{x}^{\nu}}$, where
$\dot{x}^{\mu} = dx^{\mu}/d\tau$, $d\tau = ds/c_0$. The Hamiltonian is given by
$\sum_{\mu}\,(\partial L/\partial \dot{x}^{\mu})\,\dot{x}^{\mu} - L \equiv 0$.
This implies there must be a constraint amongst the momenta, which is 
given by Eq.~(\ref{eq2}),  and acts as the effective Hamiltonian for the system \cite{Lan86},
$H_{eff} = g^{\mu\nu}\,p_{\mu}\,p_{\nu}$. Substituting in the quantization condition
$p_{\mu}\to -i\hbar\nabla_{\mu}$ yields Eq.~(\ref{eq14}) directly.

Note the most general Lagrangian allows for a term proportional 
to the Ricci scalar $R(x)=R_{\mu}^{\hspace{.5em}\mu}(x)$,
i.e. an extra term $-\xi\,R(x)\,\phi^2(x)$ in Eq.~(\ref{eq13}), which leads to a corresponding
term  $\xi\,R(x)\,\phi(x)$ in the GRKGE \cite{Bir82}. The value of  $\xi$ is not determined
by any physical principle and to date has not been experimentally measured due to
the minute effects of curvature in our solar system. The value of $\xi=0$ is called
\textit{minimal coupling} for obvious reasons. In $n$ dimensions the value of
$\xi(n) = 1/4\,(n-2)/(n-1)$, which assumes the value of $\xi=1/6$ in four dimensions,
is called \textit{conformal coupling}. For massless particles under conformal coupling
the GRKGE is invariant under the simultaneous conformal transformation 
of the metric $g_{\mu\nu}(x)\to f^2(x)\,g_{\mu\nu}(x)$ and field transformation
$\phi(x)\to f^{-1}(x)\,\phi(x)$ for an arbitrary function $f(x)$. 
In this paper we concern ourselves with regions
of spacetime devoid of matter so that Einstein's equations reduce to $R_{\mu\nu}=0$ and
consequently $R(x)=0$.

\subsection{Spin 1/2 field}
\label{sec3.05}
For the case of spin $1/2$ we have  
$\Sigma^{ab} = 1/4 [\,\gamma^a,\,\gamma^b\,]=-i/2 \,\sigma^{ab}$, 
$\partial_a \psi \to e_a^{\hspace{.5em}\mu}\nabla_{\mu} \psi$,
and we obtain
\begin{subequations}
\begin{eqnarray}
 \label{eq15}
\mathcal{L}(x) &=& \frac{i}{2}\, \left( \bar{\psi} \gamma^a \partial_a \psi 
            - (\partial_a \bar{\psi}) \gamma^a  \psi - \frac{m c_0}{\hbar}\bar{\psi} \psi \right) \nonumber \\
            & \to & (-g)^{1/2} \, \frac{i}{2}\, 
                                  \left(  \bar{\psi} \gamma^a e_a^{\hspace{.5em}\mu} \, \nabla_{\mu} \psi 
                                         -  e_a^{\hspace{.5em}\mu} \, (\nabla_{\mu} \bar{\psi}) \gamma^a  \psi 
                                         - \frac{m c_0}{\hbar}\bar{\psi} \psi 
                                  \right)  \label{eq15a} \\
            & \equiv& (-g)^{1/2}  \,\frac{i}{2}\, 
                                  \left(  \bar{\psi} \gamma^{\mu} \, \nabla_{\mu} \psi 
                                         -  (\nabla_{\mu} \bar{\psi}) \gamma^{\mu}  \psi 
                                         - \frac{m c_0}{\hbar}\bar{\psi} \psi 
                                  \right)  \label{eq15b}
\end{eqnarray}
\end{subequations}
In the above we have defined the curved spacetime counterparts to the Dirac matrices 
as $\gamma^{\mu}(x) =  \gamma^{a} \, e_a^{\hspace{.5em}\mu}(x)$ which satisfy the Clifford algebra
$\{ \gamma^{\mu}(x), \gamma^{\nu}(x) \} = 2 g^{\mu\nu}(x)$. Variation of  Eq.~(\ref{eq15b})
with respect to $\bar{\psi}(x)=\psi^{\dagger}(x)\,\gamma^0(x)$ 
yields the general relativistic Dirac equation (GRDE) for spin $1/2$ particles 
\begin{equation}
 \label{eq16}
  \left( i \, \gamma^{\mu}(x) \, \nabla_{\mu} - \frac{m c_0}{\hbar} \right) \psi(x) = 0.
\end{equation}
Note that no geometric curvature term with a dimensionless coefficient can be added
to the Lagrangian and thus to the GRDE \cite{Bir82}.

\subsection{Comment on the Pauli-Schr\"{o}dinger equation}
\label{sec3.075}
In flat spacetime, the Dirac wavefunction also satisfies the Klein-Gordon equation due to the factorization 
$\bigl( i\gamma^a\partial_a + m c_0/\hbar \bigr)\,\bigl( i\gamma^a\partial_a - m c_0/\hbar \bigr)\psi=0$
which implies $\bigl[\eta^{ab}\partial_a\partial_b +  (m c_0/\hbar)^2   \bigr]\psi=0$. However, in the 
presence of a gravitational field, this is no longer the case \cite{Cha84}. Using Eq.~(\ref{eq16}),
the analogous calculation produces
$\bigl( i\gamma^{\mu}\nabla_{\mu} + m c_0/\hbar \bigr)\,\bigl( i\gamma^{\nu}\nabla_{\nu} - m c_0/\hbar \bigr)\psi=0$ 
$\Rightarrow \bigl[\gamma^{\mu}\gamma^{\nu}\nabla_{\mu}\nabla_{\nu} +  (m c_0/\hbar)^2 \big]\psi=0$.
This can be written as 
$\bigl[g^{\mu\nu}\nabla_{\mu}\nabla_{\nu} -1/2 \sigma^{\mu\nu} K_{\mu\nu} +  (m c_0/\hbar)^2 \big]\psi=0$,
where 
$K_{\mu\nu} \equiv 1/2(\nabla_{\nu}\,\nabla_{\mu} -\nabla_{\mu}\,\nabla_{\nu})$ 
$=\partial_{\nu}\Omega_{\mu} -  \partial_{\mu}\Omega_{\nu} + [\Omega_{\nu},\Omega_{\mu}]$
is called the \textit{spin curvature} in analogy to the Riemann curvature \cite{note2}. The relation between
the spin curvature and the Riemann curvature is given by (see appendix of \cite{Cha84})
$K_{\mu\nu} = 1/4 R_{\mu\nu\alpha\beta}\sigma^{\alpha\beta}$ and
$R\textbf{I} = R_{\mu\nu\alpha\beta}\sigma^{\mu\nu}\sigma^{\alpha\beta}= -2K_{\mu\nu}\sigma^{\mu\nu}$
where $R$ is the Ricci scalar and $\textbf{I}$ is the 
$4\times 4$ unit matrix. Using the trace of Einstein's equations 
$R_{\mu\nu} - 1/2 g_{\mu\nu} R = -8\pi G/c_0^2 T_{\mu\nu}$ we obtain
\begin{equation}
\label{eq16.5}
\left[ g^{\mu\nu}\nabla_{\mu}\nabla_{\nu} + \frac{2\pi G}{c_0^2}\,T +  (m c_0/\hbar)^2 \right]\psi = 0,
\end{equation}
where $T=T_{\mu}^{\hspace{.5em}\mu}$ is the trace of the energy-momentum tensor.

%In regions of space for which $T=0$,  and Eq.~(\ref{eq16.5}) appears formally similar to the
%GRKE, Eq.~(\ref{eq14}). However, there is an important difference.  
Eq.~(\ref{eq16.5}) is the
generally covariant extension of the Pauli-Schr\"{o}dinger equation  \cite{note3} and
describes spin $1/2$ particles \textit{in a gravitational field}. As such, the covariant derivative
in  Eq.~(\ref{eq16.5}) is given by  Eq.~(\ref{eq12a}) involving the Fock-Ivanenko coefficients.
On the other hand,  Eq.~(\ref{eq14}) is the generally covariant extension of the Klein-Gordon
equation for spin $0$ particles, with no $\Omega_{\mu}$ terms. 
If we expand the spin covariant
derivatives in Eq.~(\ref{eq16.5}) using Eq.~(\ref{eq12a}) we have 
\begin{eqnarray}
\label{eq16.75}
\left[\right. 
g^{\mu\nu}\partial_{\mu}\partial_{\nu} &+& (m c_0/\hbar)^2\left.\right]\,\psi  
+ (2\pi G/c_0^2)^2\,T \,\psi 
\nonumber \\
&+& g^{\mu\nu} \big( 2\,\Omega_{\mu}\partial_{\nu} + (\partial_{\mu}\Omega_{\nu})  
+ \Omega_{\mu}\,\Omega_{\nu} \big)\psi = 0.
\end{eqnarray}
In flat spacetime and in regions where $T=0$,
the tetrads are constant unit vectors and $g^{\mu\nu}\to\eta^{\mu\nu}$ in Cartesian coordinates.
Hence, the Fock-Ivanenko coefficients are zero. The remaining terms in
the square brackets above constitute the usual Minkowski Klein Gordon equation
for the spinor field $\psi(x)$. Similarly, for the scalar wave equation
Eq.~(\ref{eq14}) in flat spacetime in Cartesian coordinates, the 
affine connection $\Gamma^{\sigma}_{\hspace{.5em}\mu\nu}$ is also zero,
and so again we recover the Minkowski Klein Gordon equation, for the scalar field $\phi(x)$.
The point to be made here is that in
flat spacetime the Dirac wave function (which describes spin $1/2$ particles) is also a solution of the 
Klein-Gordon equation (which describes spin $0$ particles) by the virtue of there being 
\textit{no gravitational field} \cite{Cha84}. In curved spacetime, this property no longer holds. In
the presence of a gravitational field, the solution to the generally covariant Dirac equation is a solution
of the generally covariant Pauli-Schr\"{o}dinger equation.

\subsection{Massless spin 1 field}
\label{sec3.1}
In Minkowski spacetime the electromagnetic (massless, spin $1$) field $F^{ab}$ is described by the
Lagrangian $\mathcal{L} = -1/4\,F^{ab}\,F_{ab}$. where 
$F_{ab}=\partial_b A_a - \partial_a A_b$ in terms of the vector potential $A_{a}$. The gauge
freedom of the vector potential prevents a straightforward quantization of the theory and
leads to the introduction of gauge-fixing terms $\mathcal{L}_G = -1/2\,\zeta^{-1}\, (\partial_a A^a)^2$
where $\zeta$ is a parameter determining the choice of the gauge. In the \textit{Feynman gauge} 
$\zeta=1$, variation of the combined action $\mathcal{L}+\mathcal{L}_G$ 
with respect to the vector potential leads to the wave equations 
$\Box A_c \equiv \eta_{ab}\,\partial_a\,\partial_b\,A_c=0$ for each component of the vector potential.

In curved spacetime the electromagnetic field takes on the same form as 
in Minkowski spacetime due to the cancellation of the connection terms,
$F_{\mu\nu} = A_{\mu;\nu} -A_{\nu;\mu} =  
\partial_{\nu} A_{\mu} - \partial_{\nu} A_{\nu}$.
To generalize the Maxwell field to curved spacetime 
we make the substitution $A_a\to A_{\mu}$, 
$\partial_a\to  e_a^{\hspace{.5em}\mu}\,\nabla_{\mu}$ with $\Omega_{\mu}(x)$
given by Eq.~(\ref{eq12c}) with 
$\left[\Sigma_{ab}\right]_c^{\hspace{.5em}d} = 
\delta_a^{\hspace{.5em}d} \eta_{bc} - \delta_b^{\hspace{.5em}d} \eta_{ac}$ 
corresponding to the $(1/2,1/2)$ representation of the Lorentz group.
The action takes the form of 
$\mathcal{L} = -1/4\,F^{\mu\nu}\,F_{\mu\nu}$ and correspondingly the gauge fixing term takes the form
$\mathcal{L}_G = -1/2\,\zeta^{-1}\, (\nabla_{\mu} A^{\mu})^2$. Variation of the action
$\mathcal{L}+\mathcal{L}_G$ leads to the wave equations
\begin{eqnarray}
\label{eq17}
A_{\mu;\nu}^{\hspace{1em}\nu} &+& R_{\mu}^{\hspace{.5em}\nu}\,A_{\nu} 
      - (1 - \zeta^{-1})\,A_{\nu;\hspace{.5em}\mu}^{\hspace{.5em}\nu} = 0, \nonumber \\
\Box A_{\mu} &=&  0, 
      \qquad \textrm{Feynman gauge and} \, R_{\mu\nu}=0
\end{eqnarray}
For our purposes, each component of the vector potential $A_{\mu}(x)$
in Eq.~(\ref{eq17}) satisfies the massless GRKGE, 
so that it suffices for us to examine the spin $0$ GRKGE Eq.~(\ref{eq14}).

\section{Plane Wave-like solutions for the GRDE}
\label{sec4}
We are now interested in finding solutions to the GRDE of the form Eq.~(\ref{eq3})
which we write as
$\psi(\bm{r},t) = A(\bm{r},t) \, e^{i\,S(\bm{r},t)/\hbar}$ where $u$ is a Dirac spinor.
Writing out  Eq.~(\ref{eq16}) and using  Eq.~(\ref{eq12b}) the GRDE is explicitly
given as
\begin{equation}
 \label{eq18}
  \left[ i \gamma^c e_c^{\hspace{.5em}{\mu}} \left( \partial_{\mu} 
              -\frac{i}{4} \sigma^{ab} e_a^{\hspace{.5em}{\nu}} 
                e_{b\nu;\mu} \right) - \frac{m c_0}{\hbar} \right] \psi(x) = 0.
\end{equation}
Using the explicit representation of the Minkowski spacetime Dirac matrices $\gamma^a$ given
in the Appendix \ref{appendixA}, substituting in the static metric  Eq.~(\ref{eq6}) 
and dividing the resulting equation by $\Phi(\bm{r})$ yields after some algebra \cite{note4}
\begin{equation}
 \label{eq19}
  \left[ i\left(\, n(\bm{r}) \,\gamma^0 \partial_0 + \bm{\gamma}\cdot\bm{\nabla} 
        +\frac{3}{4}\,\bm{\gamma}\cdot\bm{\nabla}\zeta(x)\,\right) - k_c(\bm{r})
  \right] \psi(x) = 0.
\end{equation} 
In Eq.~(\ref{eq19}) we  defined
$\bm{\nabla}=\{\partial/\partial x, \partial/\partial y, \partial/\partial z\}$ 
as the spatial gradient in the general coordinate frame and $\bm{\gamma} = \{\gamma^1,\gamma^2,\gamma^3\}$
the spatial components of the constant Dirac matrices in the local Lorentz frame.
The quantity $\zeta(x)$ is
given by $\zeta(x) \equiv \ln[\Omega(\bm{r})/\Phi^2(\bm{r})]$.

The middle term in Eq.~(\ref{eq19})  $ \frac{3}{4}\,\bm{\gamma}\cdot\bm{\nabla}\zeta(x)$
can be removed by defining the wavefunction as  $\psi({\bm{r}},t) = f(\bm{r})\, \phi(\bm{r},t)$. 
A simple calculation reveals  
$f(\bm{r}) = e^{-3/4\,\zeta(x)} = (\Phi^2(\bm{r})/\Omega(\bm{r}))^{3/4}$.
The resulting equation for $\phi(x)$ is given by
\begin{equation}
 \label{eq20}
  \left[ i \, \left(\, \gamma^0 \,\frac{n(\bm{r})}{c_0} \, \frac{\partial}{\partial t} 
         + \bm{\gamma}\cdot\bm{\nabla} \,\right) - k_c(\bm{r})
  \right] \phi(\bm{r},t) = 0.
\end{equation} 
 Eq.~(\ref{eq20}) is directly interpretable as the Dirac equation in flat spacetime
with a spatially varying index of refraction $n(\bm{r})$ and a spatially varying
Compton wavelength $\lambda_c(\bm{r})=\lambda_c^{flat}\,\Phi(\bm{r})$, 
$k_c(\bm{r})=1/\lambda_c(\bm{r})$, $\lambda_c^{flat}=\hbar/m c_0$.

We now seek a plane wave-like solution by substituting in 
Stodolsky's suggestion for the QMP in the form
\begin{equation}
\label{eq20.5}
\phi(\bm{r},t) = A \, \exp\left[\frac{i}{\hbar}\,
                 \left(\int^{\bm{r}}\,\bm{p}(\bm{x})\cdot d\bm{x}\, - E\,t\right)\right] 
\end{equation}
where $A$ is a constant Dirac spinor. Note that E is a constant of the geodesic motion given by
$E\equiv p_0 c_0 = \dot{t}\,\Omega^2(\bm{r})\,m c_0^2$ since the metric is independent of the coordinate $t$.
Separating Eq.~(\ref{eq20}) into a coupled set
of two-spinor equations we have
\begin{equation}
 \label{eq21}
\left(
\begin{array}{cc}
 n(\bm{r})\,E/c_0 - m c_0/\Phi(\bm{r}) & -\bm{\sigma}\cdot\bm{p}(\bm{r}) \\
 -\bm{\sigma}\cdot\bm{p}(\bm{r}) &  n(\bm{r})\,E/c_0 + m c_0/\Phi(\bm{r}) 
\end{array}
\right)\,
\left(
 \begin{array}{c}
    A_+ \\
    A_-
 \end{array}
\right) = 0,
\end{equation}
where $\bm{\sigma}$ are the usual $2\times 2$ Pauli spin matrices (Appendix \ref{appendixA}). 
Eq.~(\ref{eq21}) is a homogeneous set of algebraic equations and therefore
only has a nontrivial solution if the determinant of the coefficients
is zero.  Using the identity 
$(\bm{\sigma}\cdot\bm{p})^2 = \bm{p}\cdot\bm{p}\equiv\bm{p}^2$ and setting
the determinant of the coefficients equal to zero yields precisely the mass shell 
constraint in the form of Eq.~(\ref{eq8}).

To summarize, we have found an exact plane wave-like solution for the GRDE
for static metrics in isotropic coordinates to be of the form
\begin{eqnarray}
 \label{eq22}
\psi(\bm{r},t) &=& A \, \left( \frac{\Phi^2(\bm{r})}{\Omega(\bm{r})} \right)^{3/4} \, 
      \exp \left[ \frac{i}{\hbar}\,(\int^{\bm{r}}\,\bm{p}(\bm{x})\cdot d\bm{x}\, - E\,t) \right] 
      \nonumber \\
 &=& A \, \left( \frac{\Phi^2(\bm{r})}{\Omega(\bm{r})} \right)^{3/4} \,
 \exp \left(  i\, \int^{\bm{r}}\,k_{\mu}\,dx^{\mu}  \right).
\end{eqnarray}
This solution is normalizable with respect to the inner product
\begin{equation}
\label{eq22.5}
\int\,d^3\bm{r}\,\sqrt{\textrm{det}\,g_{ij}}\,\psi^{\dagger}(\bm{r},t)\,\psi(\bm{r},t),
\end{equation}
where $\textrm{det}\,g_{ij}=\Phi^{-6}(\bm{r})$  is the determinant of the spatial
portion of the metric $g_{\mu\nu}$.

Eq.~(\ref{eq22}) is one of the main results of this paper. It is an exact solution, valid
for arbitrary strength gravitational fields for both massive and massless fields.
Several authors \cite{Don86,Wil74,Fis81,Heh90,Var98}
begin with  Eq.~(\ref{eq18}) but examine the equation in the weak field limit
$e_a^{\hspace{.5em}\mu} = \delta_a^{\hspace{.5em}\mu} + 1/2\, h_a^{\hspace{.5em}\mu}$.
In some of those works, a Foldy-Wouthuysen transformation \cite{Bjo64,Gre94} is performed to 
write the GRDE in the form of an 
effective weak field Schr\"{o}dinger equation,
$i\hbar\,\partial\psi/\partial t = H\,\psi$.
Typically, one is interested in the Hamiltonian with respect to the measure $\int\,d^3\bm{r}$
(versus the measure in Eq.~(\ref{eq22.5})) so that the momentum operator can
be interpreted in the usual flat spacetime form  as a spatial gradient 
in Cartesian coordinates, $\bm{p}=-i\hbar\,\bm{\nabla}$.
In such cases, a new
wave function is defined by $\tilde{\psi}=(\textrm{det}\,g_{ij})^{1/4} \psi\equiv U\psi$,
with the corresponding Hamiltonian $\tilde{H} = U H U^{-1}$.
The focus of the above works is for the most part, the physical
interpretation of terms in the effective weak field Hamiltonian $\tilde{H}$ as 
post-Newtonian corrections to the gravitational  potential, with an emphasis on
spin-gravity coupling. 

Instead, in this work, we are interested in the solution of the GRDE Eq.~(\ref{eq18}) directly,
with emphasis on  strong gravitational fields and the issue of whether or not the phase
of the wavefunction takes the form proposed by Stodolsky. 
If one were to try force an interpretation of Eq.~(\ref{eq20}) as a Schr\"{o}dinger 
equation, the effective Hamiltonian would contain non-Hermetian terms. This fact is
well known (see \cite{Wil74,Fis81}) and arises because of our removal of terms
that involved gradients of the gravitational field, and the 
introduction of a spatially varying function $n(\bm{r})$ in front of the first order time derivative.
As stated earlier,  we interpret  Eq.~(\ref{eq20}) as the flat spacetime Dirac equation
for a particle moving in a medium with a spatially varying  effective index of refraction and
Compton wavelength. Regardless of any imposed physical interpretation, the point we wish to emphasis here
is that the GRDE admits a solution with exactly the phase suggested by Stodolsky  Eq.~(\ref{eq1}),
and a spatially varying amplitude.

\section{WKB Solution to the GRKGE}
\label{sec5}
To develop a solution for the GRKGE Eq.~(\ref{eq14}), we begin by using the identity \cite{Lau65}
$g^{\mu \nu }\Gamma _{\mu \nu }^{\lambda
}$ $=-~(-g)^{-1/2}\partial _{\mu }[(-g)^{1/2}g^{\mu \lambda }]$. Substituting this and the 
static metric  Eq.~(\ref{eq6}) into  Eq.~(\ref{eq14}) yields
\begin{equation}
\label{eq23}
\frac{n^{2}}{c_{0}^{2}}\frac{\partial ^{2}\psi }{\partial t^{2}}-\bm{%
\nabla }^{2}\psi - \bm{\nabla}\xi \cdot \bm{\nabla}\psi +k_{c}^{2}\left( \bm{r}\right) \psi =0, 
\end{equation}
where we define the quantity $\xi \equiv \ln\bigl(\Omega(\bm{r})/\Phi(\bm{r})\bigr)$.
As in the previous section, we can remove the 
term linear in $\bm{%
\nabla }\psi $  by seeking a solution of the form 
\begin{equation}
\label{eq24}
\psi \left( \bm{r},t\right) =f(\bm{r})\phi \left( \bm{r},t\right),   
\end{equation}
with $f(\bm{r})=(\Phi (\bm{r})/\Omega (\bm{r}))^{1/2}$ and $\phi
\left( \bm{r},t\right) $ satisfying 
\begin{equation}
 \label{eq25}
\frac{n^{2}}{c_{0}^{2}}\frac{\partial ^{2}\phi }{\partial t^{2}}-\bm{%
\nabla }^{2}\phi +\left[ k_{c}^{2}\left( \bm{r}\right) +\eta \left( 
\bm{r}\right) \right] \phi =0, 
\end{equation}
where
\begin{equation}
 \label{eq26}
\eta \left( \bm{r}\right) \equiv \frac{1}{2}\bm{\nabla }^{2}\xi
\left( \bm{r}\right) +\frac{1}{4}|\bm{\nabla }\xi \left( \bm{r}%
\right) \mid ^{2}. 
\end{equation}
Note Eq.~(\ref{eq25}) is essentially the flat spacetime wave equation
for a scalar particle moving in spatially varying index of refraction 
and Compton wavelength. However, there is an additional term $\eta(\bm{r})$
in the wave equation which is a direct result of the geometric cross
term $g^{\mu \nu }\Gamma _{\mu \nu }^{\lambda}$ arising from the 
product of the covariant derivatives. In the weak gravitation field
limit this term is typically dropped \cite{Don86} when seeking an approximate
solution to the wave equation. This term can also be exactly eliminated 
by transforming to  \textit{harmonic} coordinates (\cite{Wei72} p161-163) for which
the identity above becomes four coordinate conditions, namely 
$\partial _{\mu }[(-g)^{1/2}g^{\mu \lambda }]=0$. However, the use harmonic coordinates
introduces off-diagonal terms in the metric, which for the Schwarzschild case (\cite{Wei72}, p181)
take the form of $h(R) \bm{X}\cdot d\bm{X}$ where $\bm{X}$ are the new harmonic
Cartesian coordinates. In this work, we prefer to remain in isotropic coordinates,
where the metric has a simple diagonal form and its spatial portion is Euclidean conformally flat,
and remove the cross term $-\bm{\nabla}\xi \cdot \bm{\nabla}\psi$ 
arising from $g^{\mu \nu }\Gamma _{\mu \nu }^{\lambda}$ by the transformation
of the wavefunction, Eq.~(\ref{eq24}).

A direct substitution of a plane wave-like solution of the form
of  Eq.~(\ref{eq20.5}) \textit{does not} yield the mass shell constraint Eq.~(\ref{eq8}). 
This is due in part to the existence of
the term  $\eta(\bm{r})$ and the fact that the spatial Laplacian $\nabla^2$
generates imaginary terms linear in $\bm{\nabla}\cdot\bm{p}$. In the next section
we develop a WKB solution to  Eq.~(\ref{eq25}).

\subsection{The WKB expansion}
\label{WKBExpansion}

Let us look for a stationary solution to Eq.~(\ref{eq25}) of the form 
\begin{equation}
\label{eq27}
\phi \left( \bm{r},t\right) =A\exp \left[ \frac{i}{\hbar }\mathcal{S}(%
\bm{r},t)\right]   
\end{equation}
with $A$ constant and where we take the phase to be of the form 
\begin{equation}
 \label{eq28}
\mathcal{S}(\bm{r},t)=S(\bm{r})-E\,t. 
\end{equation}
Substitution of Eq.~(\ref{eq27}) and Eq.~(\ref{eq28}) into Eq.~(\ref{eq25}) yields 
\begin{equation}
\label{eq29}
\left( p^{2}-\left( \bm{\nabla }S\right) ^{2}\right) +i\hbar \bm{%
\nabla }^{2}S-\hbar ^{2}\,\eta (\bm{r})=0,  
\end{equation}
\newline
where we have used Eq.~(\ref{eq8}).

We now expand $S(\bm{r})$ in a power series expansion in $\hbar $ via 
\begin{equation}
\label{eq30}
S(\bm{r})=\sum_{n=0}^{\infty }\hbar ^{n}\,S_{n}(\bm{r}).  
\end{equation}
Substituting this into Eq.~(\ref{eq29}) yields the set of equations 
\begin{subequations}
\label{eq31}
\begin{eqnarray}
O(\hbar^{0}):\qquad \left( \bm{\nabla }S_{0}\right) ^{2}-p^{2} &=&0
\label{eq31a} \\
O(\hbar^{1}):\qquad 2\bm{\nabla }S_{0}\cdot \bm{\nabla }S_{1} &=&i%
\bm{\nabla }^{2}S_{0}  \label{eq31b} \\
O(\hbar^{2}):\qquad 2\bm{\nabla }S_{0}\cdot \bm{\nabla }S_{2} &=&i%
\bm{\nabla }^{2}S_{1}-\left( \bm{\nabla }S_{1}\right) ^{2}+\eta (%
\bm{r})  \label{eq31c} \\
O(\hbar^{n\geq 3}):\qquad 2\bm{\nabla }S_{0}\cdot \bm{\nabla }S_{n} &=&i%
\bm{\nabla }^{2}S_{n-1}-\sum_{j=1}^{n-1}\bm{\nabla }S_{j}\cdot 
\bm{\nabla }S_{n-j}  \label{eq31d}
\end{eqnarray}
\end{subequations}

\subsection{The eikonal equation}
\label{eikonaleqn}
Eq.~(\ref{eq31a}) is the dominant contribution to the phase of the wavefunction and
represents the eikonal equation. We can solve this equation for $S_{0}$ via 
\begin{equation}
 \label{eq32}
S_{0}(\bm{r})=\int^{\bm{r}}\bm{p}(\bm{x})\cdot d\bm{x} 
\end{equation}
where 
\begin{eqnarray}
\label{eq33}
p=|\bm{p}| &\equiv& \frac{E}{c_0}\,N(\bm{r}) =  
\frac{E}{c_0}\,n(\bm{r})\sqrt{1-\left(\frac{\Omega(\bm{r}) m c_0^2}{E}\right)^2} 
= \frac{E}{c_0}\,\frac{n^2(\bm{r})\,v(\bm{r})}{c_0}, \\ 
\bm{p}(\bm{r})&=&\frac{E}{c_0}\,\frac{n^{2}\bm{v}(\bm{r})}{c_{0}}.  
\end{eqnarray}
In Eq.~(\ref{eq33}) we have introduced several definitions. First, the magnitude $p$ of the
spatial momentum is obtained by rearranging  Eq.~(\ref{eq8}). Since the energy $E$ is a constant
of the motion for a static metric, the coordinate velocity $\bm{v}(\bm{r})=d\bm{r}/dt$ 
is a function of position.
Its magnitude is given by  \cite{Eva96} $v(\bm{r}) = c_0/n(\bm{r})\,\sqrt{1-(\Omega(\bm{r}) m c_0^2/E)^2}$ and
is related to $p$ as given above. Note that for light $m=0$ and $v(\bm{r})=c_0/n(\bm{r})$ which
allows the interpretation of $n(\bm{r})$ as the index of refraction for massless particles.
For massive particles $E$ is an extra degree of freedom which specifies the initial
speed of the particle. The quantity 
\begin{equation}
\label{eq34}
N(\bm{r}) =  n(\bm{r})\,\sqrt{1-\left(\frac{\Omega(\bm{r}) m c_0^2}{E}\right)^2} = 
             \frac{n^2(\bm{r})\,v(\bm{r})}{c_0},
\end{equation} 
may be interpreted as the index of refraction for massive de Broglie waves. We will return
to this point in Section \ref{sec6}.

$S_{0}(\bm{r})=$ constant are just the wave front surfaces
with normal given by 
\begin{equation}
\label{eq35}
\bm{\nabla }S_{0}(\bm{r})=\bm{p}(\bm{r})\,\,\hspace{0.25in}%
\textrm{or}\hspace{0.25in}\frac{\bm{\nabla }S_{0}(\bm{r})}{\hbar }=%
\frac{\bm{p}(\bm{r})\,}{\hbar }\equiv \frac{1}{\lambda (\bm{r})},
\end{equation}
where $\lambda(\bm{r})$ is the de Broglie wavelength of the particle.
Following Holmes \cite{Hol95},  let us
characterize the wavefronts $S_{0}(\bm{r)}$ by coordinates 
$\bm{r} = \bar{\bm{x}}(l,\alpha ,\beta )$ where $l$ is arc length
along the ray trajectory normal to surfaces of constant $S_{0}$, and $\alpha 
$ and $\beta $ are coordinates used to parameterize the wavefront surfaces
(for e.g. spherical coordinates). 
We write the unit tangent vector to the ray $d\overline{\bm{x}}/dl$ as a
vector in the direction of $\bm{\nabla }S_{0}(\bm{r})$ via 
\begin{equation}
 \label{eq36}
\frac{d\overline{\bm{x}}}{dl}=\frac{\bm{\nabla }S_{0}(\bm{r})}{%
\left| \bm{\nabla }S_{0}(\bm{r})\right| }=\frac{\bm{p}(\bm{r}%
)}{p}. 
\end{equation}
Note we can write 
\begin{equation}
 \label{eq37}
\frac{dS_{0}}{dl}=\,\frac{d\overline{\bm{x}}}{dl}\cdot \bm{\nabla }%
S_{0}\,=\left| \bm{p}(\bm{r})\right| \equiv p.\, 
\end{equation}
%%%%
%%%% I accidentally skipped the numbering here, i.e. I jump from eq37 to eq39
%%%
Thus, another way to write Eq.~(\ref{eq32}) is 
\begin{equation}
S_{0}(l,\alpha ,\beta )=\int^{l}\,p\,dl\cdot  
\end{equation}

\subsection{The transport equation}
\label{transporteqn}

We now need to solve the transport equation Eq.~(\ref{eq31b}), $2\bm{\nabla 
}S_{0}\cdot \bm{\nabla }S_{1}=i\bm{\nabla }^{2}S_{0}\cdot $ We note
for any function $F$, we can write 
\begin{equation}
 \label{eq39}
dF/dl=d\overline{\bm{x}}/dl\cdot \bm{\nabla }F=1/p\bm{\nabla }%
S_{0}\cdot \bm{\nabla }F 
\end{equation}
from the ray equation Eq.~(\ref{eq36}). Thus, substituting $\bm{\nabla }%
S_{0}\cdot \bm{\nabla }S_{1}=p\,dS_{1}/dl$ into the transport equation
gives us a first order differential equation for $S_{1}(l,\alpha ,\beta ),$%
\[
\frac{dS_{1}}{dl} = \frac{i}{2p}\bm{\nabla }^{2}S_{0},
\]
with solution 
\begin{equation}
 \label{eq40}
S_{1}(l,\alpha ,\beta )=\frac{i}{2}\int^{l}dl\frac{\bm{\nabla }^{2}S_{0}%
}{p}. 
\end{equation}
A short calculation shown in Appendix \ref{appendixB}, reveals  
\begin{equation}
\label{eq41}
\frac{\bm{\nabla }^{2}S_{0}}{p}=\frac{d(p\,J)/dl}{p\,J}, 
\end{equation}
where 
\begin{equation}
\label{eq42}
J=\left| \frac{\partial \bm{x}}{\partial (l,\alpha ,\beta )}\right| ,
\end{equation}
is the Jacobian of the transformation from the curvilinear ray coordinates $%
(l,\alpha ,\beta )$ to Cartesian coordinates. To prove Eq.~(\ref{eq42}) one
needs to show (see Appendix \ref{appendixB})  
\begin{equation}
\label{eq43}
\partial _{l}J=J\,\bm{\nabla \cdot }\left( \frac{\bm{\nabla }S_{0}}{p%
}\right) =J\,\bm{\nabla \cdot }\frac{d\overline{\bm{x}}}{dl},
\end{equation}
where $d\overline{\bm{x}}/dl$ is the unit tangent to the particle's trajectory,
normal to surfaces of constant $S_0$.
The net result is that upon substitution of Eq.~(\ref{eq41}) into Eq.~(\ref{eq40}), 
one can perform the integral to obtain 
\begin{equation}
\label{eq44}
S_{1}(\bm{r(}l,\alpha ,\beta ))=\frac{i}{2}\ln \left( \frac{p(\bm{r}%
)J(\bm{r})}{p(\bm{r}_{0})J(\bm{r}_{0})}\right) \equiv \frac{i}{2}%
\mu (\bm{r})  
\end{equation}
where we define $\mu (\bm{r})\equiv \ln \left( \frac{p(\bm{r})J(%
\bm{r})}{p(\bm{r}_{0})J(\bm{r}_{\bm{0}})}\right) $, and $%
\bm{r}_{\bm{0}}=\bm{r}|_{l=0}$. To the lowest order correction 
$O(\hbar^0)$ in the phase and amplitude, we
have found the WKB approximate solution 
\begin{eqnarray}
\label{eq45}  % first order solution
\psi _{1}\left( \bm{r},t\right) &=& \sqrt{\frac{\Phi (\bm{r})}{\Omega (\bm{r})}}\,
\phi _{1}\left( \bm{r},t\right) \nonumber \\ 
&=&A\sqrt{\frac{\Phi (\bm{%
r})}{\Omega (\bm{r})}}\sqrt{\frac{p(\bm{r}_{0})J(\bm{r}_{0})}{p(%
\bm{r})J(\bm{r})}}\exp \left[ \frac{i}{\hbar}\left( \int^{%
\bm{r}}\bm{p}(\bm{x})\cdot d\bm{x}-E\,t\right) \right] ,  
\end{eqnarray}
where the subscript $1$ on $\psi _{1}\left( \bm{r},t\right) $ indicates
that we have carried out the WKB expansion to $S_{n=1}.$ Note that Eq.~(\ref{eq45}) 
does take into account, to lowest order, the term $\xi \bm{(r)}%
\equiv \ln (\Omega(\bm{r})\Phi^{-1}(\bm{r})$ which arises from
the covariant derivative term $g^{\mu \nu }\Gamma _{\mu \nu }^{\lambda }.$
What has been left out are higher order terms involving $\bm{\nabla }%
^{2}\xi \left( \bm{r}\right) $, and $\bm{\nabla }\xi \left( \bm{r%
}\right) $ in $\eta \left( \bm{r}\right) $, which come from the quantum
corrections terms, $S_{n\geq 2}$ (which we deal with next). Note that,
already this solution is valid for strong gravitational fields (as opposed to
only weak fields as considered by Donoghue and Holstein \cite{Don86}) with the
restriction that $p^{2}\gg \hbar^{2}\eta \left( \bm{r}\right)$, corresponding to 
\begin{equation}
\label{eq46}
\lambda(\bm{r}) \,
\left\{ |\bm{\nabla}\xi(\bm{r})|,\,\left|\sqrt{\bm{\nabla}^{2}\xi(\bm{r})}\right|\right\} \ll 1.
\end{equation}
Recall that 
$\xi(\bm{r})\equiv \ln \bigl(\Omega(\bm{r})\Phi^{-1}(\bm{r})\bigr)$  
so that the above condition can still possibly hold for
reasonable distances close to horizon of a black hole, say. We investigate this in
the next section.

\subsection{Estimation of terms in Schwarzschild metric}

If we had defined $\phi \left( \bm{r},t\right) =u\left( \bm{r}\right) 
\exp( -i E t/\hbar) $ in our wave equation Eq.~(\ref{eq25}) 
the result would have been an equation of the Helmholz form
\begin{equation}
\label{eq47}
\bm{\nabla }^{2}u(\bm{r})+k^{2}(\bm{r})\left( 1-\frac{\eta (%
\bm{r})}{k^{2}(\bm{r})}\right) u(\bm{r})=0 
\end{equation}
where $p(\bm{r})=\hbar \,k(\bm{r}),\,k(\bm{r})=k_{0}\,N,$ with $%
k_{0}=\omega _{0}/c_{0}=E/\hbar c_{0}.$ Therefore we want to consider the
order of magnitude of the terms $|\bm{\nabla}\xi(\bm{r})
|^{2}/k^{2}(\bm{r})$ and $|\bm{\nabla}^{2}\xi(\bm{r})
|^{2}/k^{2}(\bm{r})$ with $\xi(\bm{r})\equiv \ln \bigl(\Omega (\bm{r})\Phi^{-1}\bm{(r)\bigr)}.$

For the Schwarzschild metric in isotropic coordinates we have 
\begin{eqnarray}
\Omega  &=&\frac{1-1/\rho }{1+1/\rho },\,\,\Phi =\frac{1}{\left( 1+1/\rho
\right) ^{2}},\,\,\,\xi =\ln (1-\frac{1}{\rho ^{2}}\bm{)},\,\,\, 
\nonumber \\
\,n &=&\frac{\left( 1+1/\rho \right) ^{3}}{\left( 1-1/\rho \right) }%
,\,\,\,\,\,k=k_{0}n\sqrt{1-\frac{\Omega ^{2}}{E^{\prime 2}}}  \label{eq48}
\end{eqnarray}
where $E^{\prime 2}\equiv E/(mc_{0}^{2})$ and $\rho =r/r_{s},$ with 
$r_{s}= GM/c_{0}^{2}=0.74\,M/M_{\odot }$ km 
(the the gravitational radius). 

For massless particles, $E^{\prime }\rightarrow \infty $, so that $k=k_{0}n$ and
we obtain 
\begin{equation}
\label{eq49}
\textrm{massless}:\frac{|\bm{\nabla }\xi (\bm{r})|^{2}}{k^{2}(\bm{r%
})}=\frac{4}{\left( k_{0}r_{s}\right) ^{2}}\frac{1}{\rho ^{6}}\frac{1}{%
\left( 1+1/\rho \right) ^{8}},\,\,\frac{|\bm{\nabla }^{2}\xi (\bm{r})%
|}{k^{2}(\bm{r})}=\frac{2}{\left( k_{0}r_{s}\right) ^{2}}\frac{1}{\rho
^{4}}\frac{\left( 1+1/\rho ^{2}\right) }{\left( 1+1/\rho \right) ^{8}}.
\end{equation}

Note the above ratios are finite for all values $1\leq \rho <\infty $ (the
valid range of the isotropic scaled radius $\rho$), even though the
numerators and denominators of the left hand sides of Eq.~(\ref{eq49})
each separately diverge as $\rho\rightarrow 1$. 
From Eq.~(\ref{eq48}), 
$k(\rho)\stackrel{\rho\rightarrow \infty }{\rightarrow }k_{0}$, so that $k_{0}$ is the usual
wavenumber of the particle at spatial infinity. For ordinary wavelengths, $%
k_{0}r_{s} \gg 1$, since $r_{s}\sim $ km, so the above ratios remain
incredibly small even down to $\rho=1$. Even if we were to
consider ultra long wavelengths such that $k_{0}r_{s}\sim 1,$ the ratios in
Eq.~(\ref{eq49}) could still be made small for values of $\rho \sim 2$, i.e. $%
r=2r_{s}$, the Schwarzschild radius. In this case, the term $\eta (\bm{r})$ arising from the
covariant derivative term $g^{\mu \nu }\Gamma _{\mu \nu }^{\lambda }$ is
essentially negligible compared to $k^2(\bm{r})$ 
all the down to the Schwarzschild radius, and for all
intents and  purposes, our wave equation for massless particles is of the form 
\begin{equation}
\label{eq50}
\bm{\nabla }^{2}u(\bm{r})+k^{2}(\bm{r})\,u(\bm{r})=0.
\end{equation}
However, in the next section 
%on the quantum corrections, 
it causes no great difficulty to carry terms involving $\eta(\bm{r})$ along formally.

\subsection{Quantum corrections}
\label{qcorrections}
The next order $\hbar$ corrections to the phase and amplitude of the scalar wavefunction  
arise from Eq.~(\ref{eq31c}). 
Using Eq.~(\ref{eq39}) we can write a first order equation for $%
dS_{2}/dl$ whose solution is given by 
\begin{equation}
\label{eq51}
S_{2}(l,\alpha ,\beta )=-\int^{l}\frac{dl}{2p}
\left[ \frac{1}{2}\bm{\nabla }^{2}\mu \bm{(r)}
-\frac{1}{4}\left( \bm{\nabla }\mu(\bm{r})\right)^{2}
+\eta \left( \bm{r}\right) \right] \equiv -\mathcal{S}_{2}(l,\alpha ,\beta ). 
\end{equation}
Since $S_{2}$ is purely real, this is an $O(\hbar ^{2}\bm{)}$ correction
to the phase.

The remaining equations Eq.~(\ref{eq31d}) can be formally solved to give 
\begin{equation}
\label{eq52}
S_{n}(l,\alpha ,\beta )=\int^{l}\frac{dl}{2p}\left[ i\bm{\nabla }%
^{2}S_{n-1}-\sum_{j=1}^{n-1}\bm{\nabla }S_{j}\cdot \bm{\nabla }%
S_{n-j}\right]   
\end{equation}
where the terms in the $\left[\hspace{1em}\right] $, involve only the previously
determined quantities $\left\{ S_{1},S_{2},\ldots ,S_{n-1}\right\} .$ The first
correction to the amplitude is $O(\hbar ^{2}\bm{)}$ and is given by $%
S_{3}$ via 
\begin{equation}
\label{eq53}
S_{3}(l,\alpha ,\beta )=i\int^{l}\frac{dl}{2p}
\left[ \bm{\nabla}^{2}S_{2}-\bm{\bm{\nabla}}\mu(\bm{r})\cdot\nabla S_{2}\right] 
\equiv i\mathcal{S}_{3}(l,\alpha ,\beta ).  
\end{equation}
Putting this all together, we have to $O(\hbar ^{2}\bm{)}$ in the phase
and amplitude 
\begin{eqnarray}
\label{eq54}
\psi _{3}\left( \bm{r},t\right) =\sqrt{\frac{\Phi (\bm{r})}{\Omega (%
\bm{r})}}\phi _{3}\left( \bm{r},t\right)  &=&A\sqrt{\frac{\Phi (%
\bm{r})}{\Omega (\bm{r})}}\sqrt{\frac{p(\bm{r}_{0})J(\bm{r}_{0})}{p(\bm{r})J(\bm{r})}}\exp 
\left[ -\hbar ^{2}\mathcal{S}_{3}(\bm{r})\right]   \nonumber \\
&&\hspace{-3em}\exp \left[ \frac{i}{\hbar }\left( \int^{\bm{r}}\bm{p}(\bm{x})\cdot d\bm{x}
-E\,t-\hbar^{2}\mathcal{S}_{2}(\bm{r})\right) \right], 
\end{eqnarray}
where the subscript $3$ on $\psi _{3}\left( \bm{r},t\right) $ indicates
that we have carried out the WKB expansion to $S_{n=3}.$

Eq.~(\ref{eq54}) reveals that for spin $0$ particles the quantum phase 
\begin{equation}
\label{eq53.5}  
S/\hbar = \frac{1}{\hbar} \, \left(\int^{\bm{r}}\,\bm{p}(\bm{x})\cdot d\bm{x}\, - E\,t\right),
\end{equation}
proposed by Stodolsky is only the lowest order (in $\hbar$) approximation to the full phase.
This is in stark contrast to the phase of the Dirac wavefunction, for which the phase 
Eq.~(\ref{eq53.5}) is exact. The failure of Eq.~(\ref{eq53.5}) to be the exact phase
for spin $0$ particles is directly attributable to the presence of second order spatial derivatives
in the GRKGE.  Since by Eq.~(\ref{eq17}), each component of the massless spin $1$ field sastifies the GRKGE 
these remarks also hold for the electromagnetic field $A_{\mu}(x)$. For the spin values considered in this work,
only spin $1/2$ particles satisfying the GRDE, which contain first order spatial derivatives,
have the phase of wavefunction given \textit{exactly} by Eq.~(\ref{eq53.5}).

\section{The Optical-Mechanical Analogy}
\label{sec6}

In this section we elucidate the optical-mechanical analogy for which the path of 
a particle in a gravitational field can be considered to arise from a spatially 
varying effective index of refraction \cite{Eva96,Als98,Eva00}. 
To lowest order in $\hbar$ for spin $0,1$ particles, 
and exactly for the case of spin $1/2$ particles, this is the classical
path the quantum particle follows. 
For static metrics in isotropic coordinates the magnitude of the momentum $p$ 
of the particle is function of position via  Eq.~(\ref{eq33}), $p(\bm{r}) = (E/c_0)\,N(\bm{r})$.
Using this and the quantization conditions Eq.~(\ref{eq9}) in the mass shell condition Eq.~(\ref{eq8})
we can define the phase velocity $v_{phase}$ and group velocity $v_g$ as
\begin{subequations}
\begin{eqnarray}
\label{eq58a}
v_{phase} &=& \frac{\omega}{k} = \frac{c_0}{N(\bm{r})}, \\
v_g       &=& \frac{\partial \omega}{\partial k} = v(\bm{r})
\label{eq58b}
\end{eqnarray}
\end{subequations}
where the velocity $v(\bm{r})$ is given by \cite{Eva96}
\begin{subequations}
\begin{eqnarray}
\label{eq59a}
v(\bm{r}) &=& \frac{c_0}{n(\bm{r})} \qquad\hspace{8.5em} \mbox{\rm massless particle}, \\
\label{eq59b}
v(\bm{r}) &=& \frac{c_0}{n(\bm{r})} \sqrt{1-\left(\frac{\Omega(\bm{r}) m c_0^2}{E}\right)^2}
                                               \qquad \mbox{\rm massive particle}.
\end{eqnarray}
\end{subequations}
Eq.~(\ref{eq58a}) allows us to identify $N(\bm{r})$ defined in Eq.~(\ref{eq34}) 
as the index of refraction for massive de Broglie waves. Similarly, Eq.~(\ref{eq59a}) and the limit
$N(\bm{r})\stackrel{m \rightarrow 0}{\rightarrow } n(\bm{r})$ allows us to identify
$n(\bm{r})$ as the index of refraction for massless de Broglie waves.

We can also add weight to the assertion that $N$ is an index of refraction
by deriving a geometrical optics ray equation for $N$. We begin with our ray
equation  Eq.~(\ref{eq36}), multiply through by $p$ and differentiate both
sides with respect to the arclength $l$ to obtain 
$$
\frac{d}{dl}\left( p\frac{d\overline{\bm{x}}}{dl}\right) =\frac{d}{dl}%
\left( \bm{\nabla }S_{0}\right) =\bm{\nabla }\frac{dS_{0}}{dl}=%
\bm{\nabla }p, 
$$
where we have used  Eq.~(\ref{eq37}) in the last equality \cite{note5}. 
Substituting in  $(E/c_0)\,N$ 
for $p$ on both sides of the above equation yields the geometrical optics ray
equation for the index of refraction $N$, 
\begin{equation}
\label{eq60}
\frac{d}{dl}\left( N\frac{d\overline{\bm{x}}}{dl}\right) =\bm{\nabla }N. 
\end{equation}
Eq.~(\ref{eq60}) is also directly derivable
from the variational principle \cite{Bor80}.
\begin{equation}
\label{eq61}
\delta \,\int \,N\,dl=0.
\end{equation}
Further discussion on the interpretation of $N(\bm{r})$ as the index of refraction for
massive de Broglie waves can be found in \cite{Eva00}.

\section{The Quantum Phase for Neutrino Oscillations}
\label{sec7}

As an application, we will calculate the neutrino oscillation formula 
based on the QMP expressions calculated above, for the assumed mixing of massive 
neutrinos following Fornengo \textit{et. al.} \cite{For97}. Our interests are two-fold:
(1) an example illustrating the explicit computation of the QMP and (2) an interpretation
of the QMP in terms of an effective index of refraction.

In flat spacetime neutrinos (spin $1/2$) produced by the weak interaction process
are created in a flavor eigenstate $|\nu_{\alpha}\rangle$ which is a superposition
of mass eigenstates  $|\nu_{k}\rangle$ i.e. 
$|\nu_{\alpha}\rangle = \sum_k\,U^*_{\alpha k}\,|\nu_{k}\rangle$. Here $U$ is 
the unitary matrix which mixes the different neutrino mass fields. What actually
propagates is the mass eigenstates, whose energy $E_k$ and momentum
$\bm{p}_k$ are related by the mass shell condition 
$E_k^2 = (\bm{p}_k c_0)^2 + (m_k c_0^2)^2$, and are determined at the production spacetime point $A$.
In general,  $E_k$, $\bm{p}_k$ and $m_k$ are different for the different mass states.
In flat spacetime each of the mass eigenstates propagates as 
$|\nu_{k}(\bm{r},t)\rangle = \exp(iS_k/\hbar)|\nu_{k}\rangle$ where 
$S_k = \bm{p}_k\cdot\bm{x} - E_k\,t$. 

Neutrino oscillations occur because the different mass states propagate 
differently due to the differences in their energies and momenta. When they arrive
at a detector located at a spacetime point $B$ which detects flavor eigenstates via the
weak interaction process, they have developed a relative phase shift. 
Interference between the different mass eigenstates at $B$ produces the neutrino 
oscillations. One assumes the mass eigenstates are produced by some coherent 
process at the spacetime point $A$ and that they are detected at the same spacetime point $B$. 
The probability that the neutrino $|\nu_{e}\rangle$ produced at $A$ is detected as
$|\nu_{\mu}\rangle$ at $B$ is given by (for two generations)
$\mathcal{P}(\nu_{e}\to\nu_{\mu}) = |\langle\nu_e |\nu_{\mu}(B)\rangle|^2 = 
\sin^2\theta\,\sin^2(S_{12}/2\hbar)$ where $\theta$ is a mixing angle, 
$S_k$ are the phases acquired by the mass eigenstates, and $S_{12}=S_1-S_2$.

Fornengo \textit{et al.} use the Stodolsky expression for the QMP as given by Eq.~(\ref{eq1}), 
reasoning that this form of the phase is valid independent of the particle's spin.
From our result Eq.~(\ref{eq22}) for spin $1/2$ particles, we see that this is indeed the
correct choice, though not from their original premise. 
Since the neutrinos all begin at the spacetime point $A$ and
are detected at the spacetime point $B$, they all experience the same
amplitude change as given in  Eq.~(\ref{eq22}), so we can ignore it.
We write the phase (without $\hbar$) as
\begin{equation}
\label{eq55}
S = \int_A^B \left(E\,\frac{dt}{dr}-p_r \right)\,dr
\end{equation}
Fornengo \textit{et al.} use the following procedure. They compute
the phase $S$ for a radial light-like trajectory with the mass shell condition
given by  Eq.~(\ref{eq2}). This assumes that one is considering ultra-relativistic
neutrinos with $E_k \gg m_k c_0^2$. For radial null geodesics we obtain the
condition Eq.~(\ref{eq7a}) which gives $dt/dr = n(\bm{r})/c_0$. The radial momentum
$p_r$ is given by  Eq.~(\ref{eq33}), 
$p_r=(E_k/c_0)\,N(\bm{r}) = (E_k/c_0)\,[1-(\Omega(\bm{r}) m_k c_0^2 / E_k)^2]^{1/2}$. Noting that
$E_k/c_0$ is a constant momentum along the geodesic we have
\begin{subequations}
\label{eq56}
\begin{eqnarray}
\label{eq56a}
S_k &=& \left(\frac{E_k}{c_0}\right) \, \int^{\bm{r}_B,t_B}_{\bm{r}_A,t_A} \Bigl( n(\bm{r})-N(\bm{r}) \Bigr)\,dr \\
\label{eq56b}
\mbox{} &=&  \left(\frac{E_k}{c_0}\right) \, \int^{\bm{r}_B,t_B}_{\bm{r}_A,t_A} n(\bm{r})\,
           \left(\, 1 - \sqrt{1-\left(\frac{\Omega(\bm{r}) m_k c_0^2}{E_k}\right)^2}  \,\right)\,dr.
\end{eqnarray}
\end{subequations}
Under the condition of ultra-relativistic neutrinos, the second term under the
radical is assumed small and the square root can be expanded to first order.
For the case of the Schwarzschild metric we define
$\rho = r/r_s$ and carrying out the integral yields
\begin{equation}
\label{eq57}
S_k^{Schw} =  \left(\frac{E_k}{c_0}\right) \, \frac{r_s}{2} \, \left( \frac{m_k c_0^2}{E_k} \right)^2 \, 
\left[ |\rho_B - \rho_A)| + \left|\frac{1}{\rho_B} - \frac{1}{\rho_A} \right| + \ldots \right].
\end{equation}
This is essentially the form that Fornengo \textit{et al.} write down in their paper
(\cite{For97}, Eq.~(\ref{eq33}) ) except that here we use isotropic coordinates, and
they make the further ultra-relativistic approximation 
$E_k\approx E_0 + \mathcal{O}(m_k^2 c_0^4 / 2 E_0)$ where $E_0$ is the energy at 
spatial infinity for a massless particle. Oscillations then occur at phase shifts
proportional to 
$(\Delta m_{kj}/2 E_0)\,~|\rho_B~-~\rho_A|$ $+ \mathcal{O}(\rho^{-1})$ 
where $\Delta m_{kj} = m_k^2 - m_j^2$.

We note two points. First, the gravitational effects are implicit in
 Eq.~(\ref{eq57}) since $\rho$ is the scaled coordinate distance.
In the presence of gravity, the neutrino propagates over
the proper distance 
$L_p$ given by $L_p = \int \sqrt{g_{rr}}\,dr$ 
$= r_s\,\int (1+1/\rho)^2\,d\rho$ $= r_s\,[ \rho - 1/\rho + 2\,\ln\rho ]$.  
Second,  Eq.~(\ref{eq56a}) allows us to interpret the phase $S_k$ as calculated
in the procedure of  Fornengo \textit{et al.} as the integrated
``optical path difference''
resulting from the difference between the index of refraction for 
a massless particle $n(\bm{r})$ and a massive particle $N(\bm{r})$ of momentum $E_k/c_0$.   
Note that if the neutrino was massless, $N(\bm{r})\to n(\bm{r})$ and $S_k=0$.
The authors chose their method of calculation
over that of calculating the phase along the classical trajectory as in Ref. \cite{Bha96}
even though the final results agree.
In the later case, the classical trajectories of different massive neutrinos reaching the
detection point at the same time must start at the production point at different times.
Thus, there are initial phases for the wave functions that must be added in ``by hand.''
Fornengo's \textit{et al.} approach calculates the interference between mass eigenstates
produced at the same spacetime point $A$ and detected at the same spacetime point $B$
connected by a null geodesics. We see that their ``mixed'' approach  can be interpreted
as the accumulation of phase due to the difference between massless and massive 
de Broglie waves.

If the calculations above were repeated for either a massless spin $0$ particle or say a photon, 
the derivation would proceed the same except the mass 
would be set to zero in Eq.~(\ref{eq56b}). This would imply the lowest order 
contribution to the phase, i.e. the ``classical phase'' Eq.~(\ref{eq55}), would be
identically zero.  The next contribution to the phase would come from 
the quantum correction terms $\exp(-i\hbar\mathcal{S}_2(\bm{r}))$ in  Eq.~(\ref{eq54}) where 
$\mathcal{S}_2(\bm{r})$ is defined by~Eq.~(\ref{eq51}).

%========================================================================================
\section{Conclusion}
\label{sec8}
In this paper we have examined the proposal that the phase of the wave function for 
quantum mechanical particle in curved spactime takes the form of 
Eq.~(\ref{eq1}), as put forth by Stodolsky \cite{Sto79}.
We investigated the wave equations for spin $0,1/2,1$ particles in the background
of an arbitrary static gravitational field which can be written 
in isotropic coordinates and developed
explicit plane wave-like solutions. We found that only for the case of spin $1/2$ does
the phase take the form of Eq.~(\ref{eq1}) \textit{exactly}. 
This was directly attributable to the first order spatial derivative 
structure of the Dirac wave equation. For spin $0$ and spin $1$
particles the phase takes the form of Eq.~(\ref{eq1}) only to lowest order in $\hbar$, due
to the second order spatial derivative structure of the corresponding wave equations.
We developed a WKB solution for spin $0$ particles which is also applicable for
spin $1$ particles. We noted that in a gravitational field, the wave function for
the generally covariant extension of the Dirac equation is not necessarily in addition
a solution to the curved spacetime Klein-Gordon equation, as is
the case in flat spacetime. We find it very intriguing that in the presence of
a gravitational field, the Dirac equation continues to admit exactly a generally covariant extension
of a plane wave-like solution with the phase given by Eq.~(\ref{eq1}), 
while the Klein-Gordon and massless spin $1$ wave equations only do so to
lowest order in $\hbar$. 

For the case of spin $1/2$ particles we calculated the 
quantum mechanical phase appropriate for neutrino flavor oscillations for radial geodesics. 
For spin $0$ and spin $1$ particles the phase of the quantum wave function 
is predominantly the classical phase as given 
by  Eq.~(\ref{eq1}), with higher order quantum corrections. 
For most applications, especially those involving the solar system,
the form of the quantum mechanical wave function in curved spacetime
assumed by Stodolsky would be essentially correct (except that the amplitude would vary spatially) 
for all practical calculations.
However, for the case of massless spin $0$ and spin $1$ particles 
and radial geodesics, we showed that the classical phase 
is zero and the higher order $\hbar$ WKB phases would be the dominant contribution.

For all the wave equations
discussed in this paper we drew an analogy for the geodesic path followed by a quantum particle in 
a static gravitational field to motion in a medium with a spatially varying 
effective index of refraction. By examining
the momentum of the quantum particle, we were able to define an effective index of refraction
$n(\bm{r})$ and $N(\bm{r})$ for massless for massive de Broglie waves, respectively.  
This allows us to extend the classical optical-mechanical
analogy to the quantum regime for arbitrary static, background gravitational fields 
which can be written in isotropic coordinates.

%========================================================================================
\appendix

\setcounter{equation}{0}
\renewcommand{\theequation}{A.\arabic{equation}}
\section{Dirac Gamma Matrices in Flat Spacetime}
\label{appendixA}
In this section we adhere to the notation of the main body of the text
and use Latin indices $\{a,b,c\}$ to indicate flat spacetime indices in
the range $\{0,1,2,3\}$. Latin indices in the middle of the alphabet
$\{i,j,k\}$ refer to spatial indices $\{1,2,3\}$. The Lorentz metric
is given by $\eta_{ab}=$diagonal$\{1,-1,-1,-1\}$.

The defining relation for the $4\times 4$ Dirac gamma matrices is
\begin{equation}
\label{A1}
\{\gamma^a, \gamma^b\} = \gamma^a \, \gamma^b + \gamma^b \, \gamma^a = 2 \eta^{ab}. 
\end{equation}
$\gamma^0$ is unitary and Hermetian 
$(\gamma^0)^2 = \textbf{I}_{4\times 4} = \gamma^0\,\gamma^{0\dagger}$, while
the spatial gamma matrices $\gamma^i$ are unitary $(\gamma^i)^{-1} = (\gamma^i)^{\dagger}$
and anti-Hermetian  $(\gamma^i)^{\dagger} = -\gamma^i$, with
$(\gamma^i)^2 = -\textbf{I}_{4\times 4} =  \gamma^{i}\gamma^{i\dagger}$.   
In this paper we use the explicit representation \cite{Gre94,Man84}
\begin{equation}
\label{A2}
\gamma^0 = \left(
                 \begin{array}{cc}
                      \textbf{I} & \textbf{0} \\
                      \textbf{0} & -\textbf{I}
                 \end{array}
           \right), \qquad
\bm{\gamma} = \left(
                 \begin{array}{cc}
                      \textbf{0} & \bm{\sigma} \\
                      -\bm{\sigma} & \textbf{0}
                 \end{array}
           \right), 
\end{equation}
where $\textbf{0}$ and  $\textbf{I}$ are the $2\times 2$ zero and identity matrix
respectively, $\bm{\gamma}=\{\gamma^1, \gamma^2, \gamma^3 \}$, and 
$\bm{\sigma}=\{\sigma^1, \sigma^2, \sigma^3 \}$ are the $2\times 2$ Pauli matrices
\begin{equation}
\label{A3}
\sigma^1 = \left(
                 \begin{array}{cc}
                      0 & 1 \\
                      1 & 0
                 \end{array}
           \right), \qquad
\sigma^2 = \left(
                 \begin{array}{cc}
                    0 & -i \\
                    i & 0
                 \end{array}
           \right), \qquad
\sigma^3 = \left(
                 \begin{array}{cc}
                      1 & 0 \\
                      0 & -1
                 \end{array}
           \right).
\end{equation} 
The Pauli matrices have the property 
$\sigma^i\,\sigma^j = \delta^{ij} + i\,\epsilon^{ijk}\,\sigma^k$.
Here $\epsilon^{ijk}$ is the Levi-Civita symbol with  $\epsilon^{123}=1$ 
and anti-symmetric in all its indices. Another useful relationship is
$(\bm{\sigma}\cdot\bm{a})\,(\bm{\sigma}\cdot\bm{b}) = 
\bm{a}\cdot\bm{b} + i \bm{\sigma}\cdot(\bm{a}\times\bm{b})$ for any arbitrary pair of spatial
vectors $\bm{a}$ and $\bm{b}$.

In the derivation of the wave equation for arbitrary spin in Section~\ref{sec3}
we introduced the anti-symmetric spin matrices 
\begin{equation}
\label{A4}
\sigma^{ab} \equiv  \frac{i}{2}\, [\gamma^a,\gamma^b] = 
                    \frac{i}{2}\, ( \gamma^a\,\gamma^b - \gamma^b\,\gamma^a ). 
\end{equation}
In the representation of  Eq.~(\ref{A2})  we have
\begin{equation}
\label{A5}
\sigma^{0k} =  i\,
           \left(
                \begin{array}{cc}
                      0 & \sigma^k \\
                      \sigma^k & 0
                 \end{array}
           \right), \qquad
\sigma^{ij} = \epsilon^{ijk} \,
              \left(
                 \begin{array}{cc}
                      \sigma^k & 0 \\
                       0  & \sigma^k
                 \end{array}
           \right). 
\end{equation}

\setcounter{equation}{0}
\renewcommand{\theequation}{B.\arabic{equation}}
\section{Proof of Eq.~(\ref{eq41})}
\label{appendixB}
In this appendix we will prove  Eq.~(\ref{eq41})
\begin{equation}
\label{B0} 
\frac{\bm{\nabla }^{2}S_{0}}{p}=\frac{d\ln \left( pJ\right)}{dl}
\end{equation}
where $J$ is the Jacobian of the transformation from curvilinear coordinates $(l,\alpha,\beta)$ 
(which describe the wave fronts $S_0(\bm{r})\,$) to Cartesian coordinates. 
In order prove this relation, we must first prove the following lemma,  Eq.~(\ref{eq43}).

\subsection{Lemma: Proof of Eq.~(\ref{eq43})}

We want to prove 
\begin{equation}
\partial _{l}J=J\,\bm{\nabla \cdot }\left( \frac{\bm{\nabla }S_{0}}{p%
}\right) =J\,\bm{\nabla \cdot }\frac{d\overline{\bm{x}}}{dl}
\label{B1}
\end{equation}
which states that the logarithmic derivative of the Jacobian $J$ along a congruence of ray
trajectories is equal to the divergence of the tangent vector field of the congruence. 

A standard result proved in most relativity books (see for e.g. \cite{Lau65} p242-243,
\cite{DIn92} p93-94), is
that for any matrix $a_{ij}$, with determinant $a$ and inverse, $%
a^{ij}=A^{ji}/a$, where $A^{ij}$ is the signed cofactor of $a_{ij}$, we have 
\begin{equation}
 \label{B2}
\frac{\partial a}{\partial x^{k}}=a\,a^{ji\,}\frac{\partial a_{ij}}{\partial
x^{k}}= a\,a^{ij\,}\frac{\partial a_{ij}}{\partial x^{k}}%
\,\,\,\,\textrm{for }a_{ij}\textrm{ symmetric.} 
\end{equation}
Let 
\begin{equation}
J=\left| \frac{\partial \bm{x}}{\partial (l,\alpha ,\beta )}\right| , 
\label{B3}
\end{equation}
be the determinant of the transformation matrix from Cartesian coordinates $%
\bm{x}$, to the curvilinear ray coordinates $(l,\alpha ,\beta )$. 
Writing out Eq.~(\ref{B2}) with $a_{ij}\rightarrow J_{ij}=\partial x^{i}/\partial
x^{\prime j}$, with $x^{i}$ as Cartesian coordinates and $x^{\prime i}$ as
curvilinear coordinates yields 
\begin{equation}
\frac{\partial J}{\partial x^{^{\prime }k}}=J\,\frac{\partial x^{\prime j}}{%
\partial x^{i}}\frac{\partial }{\partial x^{\prime k}}\left( \frac{\partial
x^{i}}{\partial x^{^{\prime }j}}\right) =J\,\frac{\partial x^{\prime j}}{%
\partial x^{i}}\frac{\partial }{\partial x^{\prime j}}\left( \frac{\partial
x^{i}}{\partial x^{^{\prime }k}}\right) =J\,\frac{\partial }{\partial x^{i}}%
\left( \frac{\partial x^{i}}{\partial x^{^{\prime }k}}\right) ,  \label{B4}
\end{equation}
where in the second equality we have interchanged the order of the differentiations 
$\partial /\partial x^{\prime k}$ and $\partial /\partial x^{\prime j}$, 
and have used the chain rule in
the last equality. If we set $k=1$ with $x^{^{\prime }1}=l$, the term in
the last $\left( {}\right) $ above is just $d\bm{x}/dl$, in component
form. Thus, with $k=1$, Eq.~(\ref{B4}) is just Eq.~(\ref{B1}) in component form.

\subsection{Proof of Eq.~(\ref{eq41}):}

Using Eq.~(\ref{B1}), we want to show  
\begin{equation}
\bm{\nabla }^{2}S_{0}/p=d\ln \left( pJ\right) /dl.  \label{B5}
\end{equation}
Using the ray equation Eq.~(\ref{eq36}) in the form 
\begin{equation}
\frac{d\overline{\bm{x}}}{dl}=\frac{\bm{\nabla }S_{0}(\bm{r})}{p}
\label{B6}
\end{equation}
we can write the first equality in Eq.~(\ref{B1}) as (expanding out the divergence) 
\begin{equation}
 \label{B7}
\frac{d\ln J}{dl}=-\frac{\bm{\nabla }S_{0}}{p}\cdot \frac{\bm{\nabla 
}p}{p}+\frac{\bm{\nabla }^{2}S_{0}}{p}=-\frac{d\overline{\bm{x}}}{dl}%
\cdot \bm{\nabla }\ln p+\frac{\bm{\nabla }^{2}S_{0}}{p}=-\frac{d\ln p%
}{dl}+\frac{\bm{\nabla }^{2}S_{0}}{p}. 
\end{equation}
Solving for $\bm{\nabla }^{2}S_{0}/p$ in Eq.~(\ref{B7}) yields 
the desired result Eq.~(\ref{B5}).

%===========================================================================
%                   End of Main Body of text
%\begin{acknowledgments}
%\end{acknowledgments}

%\begin{acknowledgments}
%\acknowledgments
\flushleft{\bf Acknowledgements}

One of the authors, P.M.A. would like to thank Mario Serna and
Arunava Bhadra for many fruitful discussions.
%\end{acknowledgments}

%\newpage


\begin{thebibliography}{10}
\bibitem{note2}
\bibinfo{note}{In this section only, we follow \cite{Cha84} and use
  $\sigma^{\mu\nu} = 1/2 [\gamma^{\mu}, \gamma^{\nu}]$.}
\end{thebibliography}
\end{document}